\documentclass[aps,prd,onecolumn,showkeys,groupedaddress,showpacs,nofootinbib]{revtex4}
\usepackage{epsfig} \usepackage{amsmath} \usepackage{amsfonts}
\usepackage{amssymb} \usepackage{graphicx} \usepackage{colordvi}
\usepackage{psfrag}

\newtheorem{Definition}{Definition}[section]
\newtheorem{Proposition}{Proposition}[section]

\newfont{\gotico}{eufm10 scaled\magstephalf}
\newfont{\qvd}{msam10 scaled\magstephalf}

\def\interior{\,\hbox{\vrule depth0pt height.6pt width4pt%
\vrule depth0pt height8pt}\;\,}

\def\de#1/de#2{\frac{\partial {#1}}{\partial {#2}}}
\def\De#1/de#2{\dfrac{\partial {#1}}{\partial {#2}}}

\def\det{{\rm det}\,}

\def\dim{{\rm dim}\,}

\def\diag{{\rm diag}}

\def\J{{\cal J}\/({\cal E}\times {\cal C})}
\def\E{{\cal E}}
\def\C{{\cal C}}

\def\pr{{\it Phys. Rev.}\ }
\def\prl{{\it Phys. Rev. Lett.}\ }
\def\pl{{\it Phys. Lett.}\ }

\def\ijmp{{\it Int. Journ. Mod. Phys.}\ }

\def\cqg{{\it Class. Quantum Grav.}\ }

\def\grg{{\it Gen. Relativ. Grav.}\ }

\def\mnras{{\it Mon. Not. R. Ast. Soc.}\ }

\begin{document}
\vskip-2cm

\title{$f\/(R)$ gravity with torsion: a geometric approach within the $\cal J$-bundles framework}

\author{S. Capozziello$^{1,2}$, R. Cianci$^{3}$, C. Stornaiolo$^{2}$, S.
Vignolo$^{3}$ }

\affiliation{$~^{1}$ Dipartimento di Scienze Fisiche,
Universit\`{a} ``Federico II'' di Napoli and $^{2}$INFN Sez. di
Napoli, Compl. Univ. Monte S. Angelo Ed. N, via Cinthia, I- 80126
Napoli (Italy)}

\affiliation{$^{3}$DIPTEM Sez. Metodi e Modelli Matematici,
Universit\`a di Genova,  Piazzale Kennedy, Pad. D - 16129 Genova
(Italy)}

 \date{\today}
\begin{abstract}
We discuss the $f(R)$-theories of gravity with torsion in the
framework of  ${\cal J}$-bundles. Such an approach is particularly
useful since the components of the torsion and curvature tensors
can be chosen as fiber ${\cal J}$-coordinates on the bundles and
then the symmetries and the conservation laws of the theory can be
easily achieved. Field equations of $f(R)$-gravity are studied in
empty space and in presence of various forms of matter  as Dirac
fields, Yang--Mills fields and spin perfect fluid. Such fields
enlarge the jet bundles framework and characterize the dynamics.
Finally we give some cosmological applications and discuss the
relations between $f(R)$-gravity and scalar-tensor theories.
\end{abstract}


\maketitle

\section{Introduction}
In some previous papers \cite{CVB1,CVB2,VC},  a new geometric
approach  for Gauge Theories and General Relativity (GR), in the
tetrad--affine formulation, has been proposed. It is called the
${\cal J}$-bundles framework (from now on the ${\cal J}$-bundles).

The starting point for the construction  of such a framework is
that several field  Lagrangians have corresponding Lagrangian
densities which depend on the fields derivatives only through
suitable antisymmetric combinations. This is the case of the
Einstein--Hilbert Lagrangian which, in the tetrad--affine
formulation, depends on the antisymmetric derivatives of the
spin--connection through the curvature.

In view of this fact, the basic idea developed  in
\cite{CVB1,CVB2,VC} consists in defining a suitable quotient space
of the first jet--bundle, making equivalent two sections which
have a first order contact with respect to the exterior
differentiation (or, equivalently, with respect to the exterior
covariant differentiation), instead of the whole set of
derivatives. The resulting fiber coordinates of the so defined new
spaces are exactly the antisymmetric combinations appearing in the
Lagrangian densities.

For GR, it  has been shown that the fiber coordinates of the
quotient space can be identified with the components of the
torsion and curvature tensors and the approach results
particularly useful in the gauge treatment of gravity (see
\cite{Ivanenko} for a general discussion).

The aim of this paper is to extend the mathematical machinery
developed in \cite{VC} to the $f(R)$-theories of gravity with
torsion \cite{CCSV1} in order to recast such theories in the
${\cal J}$-bundle formalism. As we will see, such an approach is
particularly useful to put in evidence the peculiar geometric
structures of the theories, as symmetries and conservation laws.

Before starting with this program, we want to recall what are the
physical motivations to enlarge the Hilbert--Einstein approach to
more general theories.

The basic considerations are related to cosmology and quantum
field theory since several shortcomings of GR have induced to
investigate whether such a theory is the only fundamental scheme
capable of explaining the gravitational interaction (see
\cite{Copeland,Odinojiri,GRGrew} for a review). Other motivations
to modify GR come from the old issue to construct a theory capable
of  recovering the Mach principle. \cite{brans,cimento,sciama}.

One of the most fruitful approaches has been that of {\it Extended
Theories of Gravity} (ETGs)  which have become a sort of paradigm
in the study of gravitational interaction. It is based on taking
into account physically motivated corrections and enlargements of
the Hilbert--Einstein action. In particular, the $f(R)$--gravity
is the simplest extension consisting in relaxing the very
stringent hypothesis that the only gravitational action is that
linear in the Ricci scalar $R$: in $f(R)$-gravity, generic
functions of the Ricci scalar are taken into account, but the
approach can be extended to include any curvature invariant and/or
any scalar fields because, from a conceptual point of view, there
are no {\it a priori} reason to restrict the gravitational
Lagrangian to a linear function of the Ricci scalar $R$, minimally
coupled with matter \cite{francaviglia}.

However ETGs are not the  "full quantum gravity" but they are
needed as working schemes toward it.  In fact, every unification
scheme, as Superstrings, Supergravity and Grand Unified Theories,
takes into account effective actions where non--minimal couplings
or higher--order terms in the curvature invariants come out.
Specifically, this scheme has been adopted in order to deal with
the quantization on curved spacetimes and the result has been that
the interactions between quantum  fields and background geometry
(or the gravity self--interactions) yield correction terms in the
Einstein--Hilbert Lagrangian \cite{odintsov,birrell}. Moreover, it
has been realized that such corrections are inescapable, if we
want to obtain the effective action of quantum gravity on scales
closed to the Planck length \cite{vilkovisky}; finally the idea
that there are no ``exact'' law of physics but that the effective
interactions can be described by ``stochastic'' functions with
local gauge invariance  (i.e. conservation laws) has been recently
taken into serious consideration \cite{ottewill}.

A part the fundamental physics motivations, all these theories
have acquired a huge interest in early and late time cosmology due
to the fact that they ``naturally" exhibit inflationary eras able
to overcome the shortcomings of Standard Cosmological Model
\cite{starobinsky,kerner} and well fit the Dark Energy issues of
the today observed accelerated behavior \cite{noi}. The related
cosmological models seem realistic and capable, in principle, of
matching with the observations \cite{ladek,mimick}.

In all these approaches, the problem of reducing   general
theories to the Einstein standard form plus scalar fields has been
extensively treated; one can see that, through a Legendre
transformation on the metric, higher--order theories, under
suitable regularity conditions on the Lagrangian, take the form of
the Einstein one in which a scalar field (or more than one) is the
source of the gravitational field; on the other side, it has been
studied the equivalence between models with variable gravitational
coupling with the Einstein gravity through a suitable conformal
transformation. In any case, the debate on the physical meaning of
conformal transformations has not been solved up to now (see
\cite{faraoni,conformalCQG} and references therein for a
comprehensive review). Several authors claim for a true physical
difference between Jordan frame (higher-order theories and/or
variable gravitational coupling) since there are experimental and
observational evidences which point out that the Jordan frame is
suitable to  match solutions with data. Others state that the true
physical frame is the Einstein one according to the energy
theorems \cite{magnano-soko}. In any case, the discussion is open
and no definite statement has been achieved up to now. The problem
can be faced from a more general viewpoint and the Palatini
approach to gravity could be useful to this goal. The Palatini
approach   was firstly introduced and analyzed by Einstein himself
\cite{palaeinstein}. It was however called the Palatini approach
as a consequence of historical misunderstandings \cite{frafe}. The
fundamental idea at the bases of the Palatini formalism is to
consider the connection $\Gamma$, entering the definition of the
Ricci tensor, to  be independent of the metric $g$, defined on the
spacetime ${\cal M}$. The Palatini formalism for the standard
Hilbert--Einstein torsion--less theory results to be equivalent to
the purely metric theory: this follows from the fact that the
field equations for the connection fields result exactly  the same
connection $\Gamma$, firstly considered independent: it is the
Levi-Civita connection of the metric $g$. There is, consequently,
no reason to impose the Palatini variational principle in the
standard Hilbert-Einstein theory instead of the metric (Einstein)
variational principle.

The situation, however, completely changes when we consider the
case of ETGs depending on analytic functions of  curvature
invariants, as $f(R)$, or non-minimally coupled  scalar fields. In
these cases,  the Palatini and the metric variational principle
provide different field equations  \cite{ACCF}.

From a physical viewpoint, considering the metric $g$ and the
connection $\Gamma$ as independent fields is somehow equivalent to
decouple the metric structure of spacetime and its geodesic
structure (i.e. the connection is not the Levi-Civita connection
of $g$), governing respectively the chronological structure of
spacetime and the trajectories of particles, moving in it. This
decoupling enriches the geometric structure of spacetime and
generalizes the purely metric formalism. This metric-affine
structure of spacetime (here, we simply mean that a connection
$\Gamma$ and a metric $g$ are involved) is naturally translated,
by means of the same (Palatini) field equations, into a bi-metric
structure of spacetime. Beside the \textit{physical} metric $g$,
another metric $h$ has to be considered. This new metric, at least
in the $f(R)$ theories, is simply related to the connection. As a
matter of facts, the connection $\Gamma$  results to be the
Levi-Civita connection of $h$ and thus provides the geodesic
structure of spacetime.

A further ingredient to generalize this metric--affine formalism
is considering also torsion in $f(R)$-gravity. In \cite{CCSV1}, we
have discussed this issue showing that the torsion field plays a
fundamental role into dynamics with remarkable applications in
cosmology. Here we want to develop $f(R)$-gravity with torsion in
the framework of the ${\cal J}$-bundles.

The paper is organized as  follows.  In Sect. II, we briefly
sketch the main features of the metric-affine approach to
$f(R)$-gravity with torsion as discussed in \cite{CCSV1}.

In Sect. III, we review the construction of $\cal J$-bundles and
their main geometric properties. We recall the definition of the
Poincar\'e--Cartan form associated with a given Lagrangian and
derive field equation from a variational principle built on the
new space.

Sect. IV is devoted to the application of  ${\cal J}$-bundles
geometry to the tetrad--affine formulation of $f(R)$-gravity with
torsion. We derive field equations in vacuum and in presence of
matter. We study explicitly the coupling with Dirac fields,
Yang--Mills fields and spin fluids, giving some  applications to
cosmological models. Finally, we discuss the equivalence between
the $f(R)$-gravity and scalar-tensor gravity with torsion.
Conclusions are drawn in Sect.V.

\section{$f(R)$-gravity with torsion: preliminaries}

For convenience of the reader, we briefly sketch the theory
discussed in \cite{CCSV1}. In  $f(R)$-gravity  with torsion, the
dynamical fields are the  pairs $(g,\Gamma)\/$ consisting of a
pseudo--Riemannian metric $g\/$ and a metric compatible linear
connection $\Gamma\/$ on the space--time manifold ${\cal M}$. Such
a theory is based on the action functional
\begin{equation}\label{00.1}
{\cal A}\/(g,\Gamma)=\int{\sqrt{|g|}f\/(R)\,ds}
\end{equation}
where $f(R)$ is a real function, $R\/(g,\Gamma) = g^{ij}R_{ij}\/$
(with $R_{ij}:= R^h_{\;\;ihj}\/$) is the  curvature scalar
associated with the connection $\Gamma\/$ and $ds :=
dx^1\wedge\dots\wedge dx^4\/$. Throughout the paper, we use the
index notation
\begin{equation}\label{00.2}
R^h_{\;\;kij}=\de{\Gamma_{jk}^{\;\;\;h}}/de{x^i} -
\de{\Gamma_{ik}^{\;\;\;h}}/de{x^j} +
\Gamma_{ip}^{\;\;\;h}\Gamma_{jk}^{\;\;\;p} -
\Gamma_{jp}^{\;\;\;h}\Gamma_{ik}^{\;\;\;p}
\end{equation}
for the curvature tensor and
\begin{equation}\label{00.3}
\nabla_{\de /de{x^i}}\de /de{x^j} = \Gamma_{ij}^{\;\;\;h}\,\de
/de{x^h}
\end{equation}
for the connection coefficients.

As it is well known, given a  metric tensor $g_{ij}\/$, every
$g$-metric compatible connection $\Gamma\/$ can be represented as
\begin{equation}\label{00.4}
\Gamma_{ij}^{\;\;\;h} =\tilde{\Gamma}_{ij}^{\;\;\;h} -
K_{ij}^{\;\;\;h}
\end{equation}
where (in the holonomic basis $\left\{ \de /de{x^i}, dx^i
\right\}\/$) $\tilde{\Gamma}_{ij}^{\;\;\;h}\/$ denote the
coefficients  of the Levi--Civita connection associated with the
metric $g_{ij}\/$ and $K_{ij}^{\;\;\;h}\/$ indicate the components
of the contortion tensor \cite{Hehl}.  Therefore, the degrees of
freedom of the theory may be identified with the (independent)
components of the tensors $g_{ij}\/$ and $K_{ij}^{\;\;\;h}\/$ (the
contortion tensor satisfies the antisymmetry property
$K_{i}^{\;\;jh} = - K_{i}^{\;\;hj}\/$).

Variations with respect to the metric  and the connection
(contortion) give rise to the  field equations  \cite{CCSV1}
\begin{subequations}\label{00.5}
\begin{equation}\label{00.5a}
f'\/(R)R_{ij} - \frac{1}{2}f\/(R)g_{ij}=0
\end{equation}
\begin{equation}\label{00.5b}
T_{ij}^{\;\;\;h} = -
\frac{1}{2f'}\de{f'}/de{x^p}\/\left(\delta^p_i\delta^h_j -
\delta^p_j\delta^h_i\right)
\end{equation}
\end{subequations}
where $T_{ij}^{\;\;\;h}:=\Gamma_{ij}^{\;\;\;h} -
\Gamma_{ji}^{\;\;\;h}\/$ denotes the torsion tensor. The presence
of matter is embodied in the action functional \eqref{00.1} by
adding to the gravitational Lagrangian a suitable matter
Lagrangian density ${\cal L}_m\/$, namely
\begin{equation}\label{00.6}
{\cal A}\/(g,\Gamma)=\int{\left(\sqrt{|g|}f\/(R) + {\cal
L}_m\right)\,ds}
\end{equation}
In \cite{CCSV1}, we have supposed the matter Lagrangian being
independent of the connection. In this case the field equations
result to be
\begin{subequations}\label{00.7}
\begin{equation}\label{00.7a}
f'\/(R)R_{ij} - \frac{1}{2}f\/(R)g_{ij}=\Sigma_{ij}
\end{equation}
\begin{equation}\label{00.7b}
T_{ij}^{\;\;\;h} = -
\frac{1}{2f'\/(R)}\de{f'\/(R)}/de{x^p}\/\left(\delta^p_i\delta^h_j
- \delta^p_j\delta^h_i\right)
\end{equation}
\end{subequations}
where ${\displaystyle \Sigma_{ij}:= -
\frac{1}{\sqrt{|g|}}\frac{\delta{\cal L}_m}{\delta g^{ij}}\/}$
plays the role of the energy--momentum tensor.

\section{The $\cal J$-bundles formalism}
\subsection{The geometric framework}
Let ${\cal M}$ be a $4$-dimensional orientable space--time
manifold, with a metric tensor $g\/$ of signature
$\eta=(1,3)=(-1,1,1,1)\/$. Let us denote by $\E\/$ the co--frame
bundle of ${\cal M}$. Moreover, let $P\to {\cal M}$ be a principal
fiber bundle over ${\cal M}$, with structural group
$G=SO\/(1,3)\/$. We denote by $\C :=J_1\/(P)/SO\/(1,3)\/$ the
space of principal connections over $P\/$. We refer $\E\/$ and
$\C\/$ to local coordinates $x^i,e^\mu_i\/$ ($i,\mu =1,\dots,4\/$)
and $x^i,\omega_i^{\;\;\mu\nu}\/$ ($\mu < \nu\/$) respectively.

The configuration space of  the theory is the fiber product
$\E\times_{{\cal M}}\C\/$ ($\E\times\C\/$ for short) over ${\cal
M}$. The dynamical fields are (local) sections of $\E\times\C\/$,
namely pairs formed by a (local) tetrad field
$e\/(x)=e^\mu_i\/(x)\,dx^i\/$ and a principal connection $1$-form
$\omega\/(x)= \omega_i^{\;\;\mu\nu}\/(x)\,dx^i\/$. We notice that
the connection $\omega\/(x)\/$ is automatically metric--compatible
with the metric $g\/(x)=\eta_{\mu\nu}\,e^\mu\/(x)\otimes
e^\nu\/(x)\/$ ($\eta_{\mu\nu}:=\diag\/(-1,1,1,1)\/$), induced on
${\cal M}$ by the tetrad field $e^\mu\/(x)\/$ itself.

We consider the first $\cal J$-bundle ${\cal J}\/(\E\times\C)\/$
(see \cite{VC}) associated with the fibration $\E\times\C\to {\cal
M}$. It is built similarly to an ordinary ${\cal J}$-bundle, but
the first order contact between sections is calculated with
respect to exterior (or exterior covariant) differentials. $\cal
J$-bundles have been recently used to provide new geometric
formulations of gauge theories and GR
\cite{CVB1,CVB2,VC,VCB2,VCB1,VCB3,VM}.

For convenience of the reader, we briefly recall the construction
of the bundle ${\cal J}\/(\E\times\C)\/$. Let
$J_1\/(\E\times\C)\/$ be the first ${\cal J}$-bundle associated
with $\E\times\C\to {\cal M}\/$, referred to local
jet--coordinates
$x^i,e^\mu_i,\omega_i^{\;\;\mu\nu},e^\mu_{ij},\omega_{ij}^{\;\;\;\mu\nu}\/$.
We introduce on $J_1\/(\E\times\C)\/$ the following equivalence
relation. Let
$z=(x^i,e^\mu_i,\omega_i^{\;\;\mu\nu},e^\mu_{ij},\omega_{ij}^{\;\;\;\mu\nu})\/$
and
$\hat{z}=(x^i,\hat{e}^\mu_i,\hat{\omega}_i^{\;\;\mu\nu},\hat{e}^\mu_{ij},\hat{\omega}_{ij}^{\;\;\;\mu\nu})\/$
be two elements of $J_1\/(\E\times\C)\/$, having the same
projection $x\/$ over ${\cal M}$. Denoting by
$\{e^\mu\/(x),\omega^{\mu\nu}\/(x)\}\/$ and
$\{\hat{e}^\mu\/(x),\hat{\omega}^{\mu\nu}\/(x)\}\/$ two different
sections of the bundle $\E\times\C\to {\cal M}\/$, respectively
chosen among the representatives of the equivalence classes $z\/$
and $\hat z\/$, we say that $z\/$ is equivalent to $\hat z\/$ if
and only if
\begin{subequations}\label{0.0}
\begin{equation}
e^\mu\/(x) = \hat e^\mu\/(x), \qquad
\omega^{\mu\nu}\/(x)=\hat{\omega}^{\mu\nu}\/(x)
\end{equation}
and
\begin{equation}
d\/e^\mu\/(x)=  d\/\hat{e}^\mu\/(x), \qquad
D\omega^{\mu\nu}\/(x)=D\hat{\omega}^{\mu\nu}\/(x)
\end{equation}
\end{subequations}
where $D\/$ is the covariant differential  induced by the
connection. In local coordinates, it is easily seen that $z \sim
\hat{z}\/$ if and only if the following identities hold
\begin{subequations}\label{0.00}
\begin{equation}
e^\mu_i=\hat{e}^\mu_i, \qquad
\omega_i^{\;\;\mu\nu}=\hat{\omega}_i^{\;\;\mu\nu}
\end{equation}
\begin{equation}
(e^\mu_{ij}- e^\mu_{ji})=(\hat{e}^\mu_{ij}- \hat{e}^\mu_{ji}),
\qquad (\omega_{ij}^{\;\;\;\mu\nu}-\omega_{ji}^{\;\;\;\mu\nu})=
(\hat{\omega}_{ij}^{\;\;\;\mu\nu}-\hat{\omega}_{ji}^{\;\;\;\mu\nu})
\end{equation}
\end{subequations}
We denote by ${\cal J}\/(\E\times\C)\/$  the quotient space
$J_1\/(\E\times\C)/\sim\/$ and by $\rho: J_1\/(\E\times\C)\to
{\cal J}\/(\E\times\C)\/$ the corresponding canonical projection.
A system of local fiber coordinates on the bundle ${\cal
J}\/(\E\times\C)\/$ is provided by
$x^i,e^\mu_i,\omega_i^{\;\;\mu\nu},E^\mu_{ij}:=\frac{1}{2}\/\left(e^\mu_{ij}-
e^\mu_{ji}\right),\Omega_{ij}^{\;\;\;\mu\nu}:=\frac{1}{2}\/\left(\omega_{ij}^{\;\;\;\mu\nu}-\omega_{ji}^{\;\;\;\mu\nu}\right)\/$
$(i<j)\/$.

The geometry of $\cal J$-bundles has been thoroughly examined in
Refs. \cite{CVB1,CVB2,VC}. As a matter of fact, the quotient
projection $\rho\/$ endows the bundle ${\cal J}\/(\E\times\C)\/$
with most of the standard features of jet--bundles geometry ($\cal
J$-extension of sections, contact forms, $\cal J$-prolongation of
morphisms and vector fields), which are needed to implement
variational calculus on ${\cal J}\/(\E\times\C)\/$.

Referring the reader to  \cite{CVB1,CVB2,VC} for a detailed
discussion on $\cal J$-bundles geometry, the relevant fact we need
to recall here is that the components of the torsion and curvature
tensors can be chosen as fiber $\cal J$-coordinates on ${\cal
J}\/(\E\times\C)\/$. In fact, the following relations
\begin{subequations}\label{0.000}
\begin{equation}
T^\mu_{ij}=2E^\mu_{ji} +
\omega^{\;\;\mu}_{i\;\;\;\lambda}\/e^\lambda_j -
\omega^{\;\;\mu}_{j\;\;\;\lambda}\/e^\lambda_i
\end{equation}
\begin{equation}
R_{ij}^{\;\;\;\;\mu\nu} = 2\Omega_{ji}^{\;\;\;\mu\nu} +
\omega^{\;\;\mu}_{i\;\;\;\lambda}\omega_j^{\;\;\lambda\nu} -
\omega^{\;\;\mu}_{j\;\;\;\lambda}\omega_i^{\;\;\lambda\nu}
\end{equation}
\end{subequations}
can be regarded as fiber coordinate  transformations on ${\cal
J}\/(\E\times\C)\/$, allowing to refer the bundle ${\cal
J}\/(\E\times\C)\/$ to local coordinates
$x^i,e^\mu_i,\omega_i^{\;\;\mu\nu},T^\mu_{ij}\/$ $(i<j)$,
$R_{ij}^{\;\;\;\;\mu\nu}\/$ $(i<j, \mu <\nu)\/$. In such
coordinates, local sections $\gamma : {\cal M}\to{\cal
J}\/(\E\times\C)\/$ are expressed as
\begin{equation}\label{0.1}
\gamma : x \to
(x^i,e^\mu_i\/(x),\omega_i^{\;\;\mu\nu}\/(x),T^\mu_{ij}\/(x),R_{ij}^{\;\;\;\;\mu\nu}\/(x))
\end{equation}
In particular, a section $\gamma\/$ is said holonomic if it is the
$\cal J$-extension $\gamma ={\cal J}\sigma\/$ of a section $\sigma
:{\cal M}\to \E\times\C\/$. In local coordinates, a section is
holonomic if it satisfies the relations  \cite{VC}
\begin{subequations}\label{0.2}
\begin{equation}
T^\mu_{ij}\/(x) = \de{e^\mu_j}\/(x)/de{x^i} -
\de{e^\mu_i}\/(x)/de{x^j} +
\omega^{\;\;\mu}_{i\;\;\;\lambda}\/(x)e^\lambda_j\/(x) -
\omega^{\;\;\mu}_{j\;\;\;\lambda}\/(x)e^\lambda_i\/(x)
\end{equation}
\begin{equation}
R_{ij}^{\;\;\;\;\mu\nu}\/(x) =
\de{\omega_{j}^{\;\;\mu\nu}\/(x)}/de{x^i} -
\de{\omega_{i}^{\;\;\mu\nu}\/(x)}/de{x^j} +
\omega^{\;\;\mu}_{i\;\;\;\lambda}\/(x)\omega_j^{\;\;\lambda\nu}\/(x)
-
\omega^{\;\;\mu}_{j\;\;\;\lambda}\/(x)\omega_i^{\;\;\lambda\nu}\/(x)
\end{equation}
\end{subequations}
namely if the quantities $T^\mu_{ij}\/(x)\/$ and
$R_{ij}^{\;\;\;\;\mu\nu}\/(x)\/$  are  the components of  torsion
and curvature tensors associated with the tetrad $e^\mu_i\/(x)\/$
and the connection $\omega_i^{\;\;\mu\nu}\/(x)\/$, in turn,
represents the section $\sigma\/$.

We also recall that the bundle ${\cal J}\/(\E\times\C)\/$ is
endowed with a suitable contact bundle. The latter is locally
spanned by the following $2$-forms
\begin{subequations}\label{0.3}
\begin{equation}\label{0.3a}
\theta^\mu = de^\mu_i \wedge dx^i + E^\mu_{ij}\,dx^i \wedge dx^j
\end{equation}
\begin{equation}\label{0.3b}
\theta^{\mu\nu} = d\omega_i^{\;\;\mu\nu}\wedge dx^i +
\Omega_{ij}^{\;\;\;\mu\nu}\,dx^i \wedge dx^j
\end{equation}
\end{subequations}
It is easily seen that a section $\gamma :{\cal M}\to{\cal
J}\/(\E\times\C)\/$ is holonomic if and only if it satisfies the
condition $\gamma^*(\theta^\mu)=\gamma^*(\theta^{\mu\nu})=0\/$
$\forall \mu,\nu =1,\ldots,4\/$. Moreover, in the local
coordinates $\{x,e,\omega,T,R\}$, the $2$-forms \eqref{0.3} can be
expressed as
\begin{equation}\label{0.4}
\theta^\mu = \tau^\mu - T^\mu \qquad{\rm and} \qquad
\theta^{\mu\nu}= \rho^{\mu\nu} - R^{\mu\nu}
\end{equation}
being $\tau^\mu = de^\mu_i \wedge dx^i +
\omega_{j\;\;\;\nu}^{\;\;\mu}e^\nu_i\,dx^j \wedge dx^i\/$, $T^\mu
= \frac{1}{2}T^\mu_{ij}\,dx^i \wedge dx^j\/$, $\rho^{\mu\nu}=
d\omega_i^{\;\;\mu\nu}\wedge dx^i +
\frac{1}{2}\left(\omega_{j\;\;\;\lambda}^{\;\;\mu}\omega_i^{\;\;\lambda\nu}
-
\omega_{j\;\;\;\lambda}^{\;\;\nu}\omega_i^{\;\;\lambda\mu}\right)\,dx^j
\wedge dx^i\/$ and
$R^{\mu\nu}=\frac{1}{2}R_{ij}^{\;\;\;\;\mu\nu}\,dx^i\wedge dx^j\/$

\subsection{The field equations}
We call a Lagrangian on $\J\/$ any horizontal $4$-form, locally
expressed as
\begin{equation}\label{1.0}
L={\cal
L}\/(x^i,e^\mu_i,\omega_i^{\;\;\mu\nu},T^\mu_{ij},R_{ij}^{\;\;\;\;\mu\nu})\,ds
\end{equation}
Associated with any of such a Lagrangian there is a corresponding
Poincar\'e--Cartan $4$-form, having local expression  (see
\cite{VC})
\begin{equation}\label{1.1}
\Theta = {\cal L}\, ds - \frac{1}{2}\de{\cal
L}/de{T_{hk}^\alpha}\,\theta^\alpha\wedge ds_{hk} -
\frac{1}{4}\de{\cal
L}/de{R_{hk}^{\;\;\;\;\alpha\beta}}\,\theta^{\alpha\beta}\wedge
ds_{hk}
\end{equation}
where ${\displaystyle ds_{hk}:=\de /de{x^h}\interior\de
/de{x^k}\interior ds\/}$. Taking the identities $dx^t\wedge
ds_{ij}=-\delta^t_j\,ds_i + \delta^t_i\,ds_j\/$ and $dx^p\wedge
dx^t\wedge ds_{ij}=-\left(\delta^p_i\delta^t_j -
\delta^p_j\delta^t_i\right)\,ds\/$  into account, it is easily
seen that the $4$-form \eqref{1.1} may be expressed as
\begin{equation}\label{1.2}
\begin{split}
\Theta = {\cal L}\,ds - \de{\cal L}/de{T_{hk}^\alpha}\left(de^\alpha_h \wedge ds_k - \omega_{h\;\;\;\nu}^{\;\;\alpha}e^\nu_k\,ds + \frac{1}{2}T_{hk}^\alpha\,ds\right)\\
-\frac{1}{2}\de{\cal
L}/de{R_{hk}^{\;\;\;\;\alpha\beta}}\left(d\omega_h^{\;\;\alpha\beta}\wedge
ds_k -
\omega_{h\;\;\;\lambda}^{\;\;\alpha}\omega_k^{\;\;\lambda\beta}\,ds
+ \frac{1}{2}R_{hk}^{\;\;\;\;\alpha\beta}\,ds\right)
\end{split}
\end{equation}
The field equations  are derived from the variational principle
\begin{equation}\label{1.2bis}
{\cal A}\/(\sigma)=\int{\cal J}\sigma^*\/(\Theta)=\int{\cal
J}\sigma^*\/({\cal L}\,ds)
\end{equation}
where $\sigma:{\cal M}\to \E\times\C\/$ denotes any section and
${\cal J}\sigma:{\cal M}\to \J\/$ its $\cal J$-extension
satisfying eqs. \eqref{0.2}.

Referring the reader to \cite{VC}  for a  detailed discussion, we
recall here that the corresponding Euler--Lagrange equations can
be expressed as
\begin{equation}\label{1.4}
{\cal J}\sigma^*\left({\cal J}\/(X)\interior d\Theta\right)=0
\end{equation}
for all $\cal J$-prolongable vector fields X on $\E\times\C\/$.
Moreover, we notice that the expression of $\cal J$-prolongable
vector fields and their $\cal J$-prolongations, involved in eq.
\eqref{1.4}, is not needed here. In order to make explicit eq.
\eqref{1.4}, we calculate the differential of the form
\eqref{1.2}, that is
\begin{equation}\label{1.3}
\begin{split}
d\Theta = d{\cal L}\wedge ds - d\left(\de{\cal L}/de{T_{hk}^\alpha}\right)\wedge\left(de^\alpha_h \wedge ds_k - \omega_{h\;\;\;\nu}^{\;\;\alpha}e^\nu_k\,ds + \frac{1}{2}T_{hk}^\alpha\,ds\right)\\
-\de{\cal L}/de{T_{hk}^\alpha}\/\left(-e^\nu_k\,d\omega_{h\;\;\;\nu}^{\;\;\alpha}\wedge ds - \omega_{h\;\;\;\nu}^{\;\;\alpha}\,de^\nu_k \wedge ds + \frac{1}{2}dT_{hk}^\alpha \wedge ds\right)\\
- \frac{1}{2}d\left(\de{\cal L}/de{R_{hk}^{\;\;\;\;\alpha\beta}}\right)\wedge\left(d\omega_h^{\;\;\alpha\beta}\wedge ds_k - \omega_{h\;\;\;\lambda}^{\;\;\alpha}\omega_k^{\;\;\lambda\beta}\,ds + \frac{1}{2}R_{hk}^{\;\;\;\;\alpha\beta}\,ds\right) \\
- \frac{1}{2}\de{\cal L}/de{R_{hk}^{\;\;\;\;\alpha\beta}}\left( -2\omega_{h\;\;\;\lambda}^{\;\;\alpha}d\,\omega_k^{\;\;\lambda\beta}\wedge ds + \frac{1}{2}\,dR_{hk}^{\;\;\;\;\alpha\beta}\wedge ds\right)=\\
\de{\cal L}/de{e^\mu_q}\,de^\mu_q\wedge ds + \frac{1}{2}\de{\cal L}/de{\omega_h^{\;\;\alpha\beta}}\,d\omega_h^{\;\;\alpha\beta}\wedge ds\\
 - d\left(\de{\cal L}/de{T_{hk}^\alpha}\right)\wedge\left(de^\alpha_h \wedge ds_k - \omega_{h\;\;\;\nu}^{\;\;\alpha}e^\nu_k\,ds + \frac{1}{2}T_{hk}^\alpha\,ds\right)\\
+ \de{\cal L}/de{T_{hk}^\alpha}\/\left(e^\nu_k\,d\omega_{h\;\;\;\nu}^{\;\;\alpha}\wedge ds + \omega_{h\;\;\;\nu}^{\;\;\alpha}\,de^\nu_k \wedge ds  \right) + \de{\cal L}/de{R_{hk}^{\;\;\;\;\alpha\beta}}\,\omega_{h\;\;\;\lambda}^{\;\;\alpha}d\,\omega_k^{\;\;\lambda\beta}\wedge ds\\
- \frac{1}{2}d\left(\de{\cal
L}/de{R_{hk}^{\;\;\;\;\alpha\beta}}\right)\wedge\left(d\omega_h^{\;\;\alpha\beta}\wedge
ds_k -
\omega_{h\;\;\;\lambda}^{\;\;\alpha}\omega_k^{\;\;\lambda\beta}\,ds
+ \frac{1}{2}R_{hk}^{\;\;\;\;\alpha\beta}\,ds\right)
\end{split}
\end{equation}
Choosing infinitesimal deformations $X\/$ of the special form
\begin{equation}\label{1.5}
X=G^\mu_q\/(x)\,\de /de{e^\mu_q} +
\frac{1}{2}G^{\mu\nu}_q\/(x)\,\de /de{\omega_q^{\;\;\mu\nu}}
\end{equation}
we have then  \cite{VC})
\begin{equation}\label{1.6}
\begin{split}
{\cal J}\/(X)\interior d\Theta = \left[\de{\cal L}/de{e^\mu_q}\,ds + d\left(\de{\cal L}/de{T_{qk}^\mu}\right)\wedge ds_k + \de{\cal L}/de{T_{hq}^\alpha}\omega_{h\;\;\;\mu}^{\;\;\alpha}\,ds \right]G^\mu_q\\
+\left[ \frac{1}{2}\de{\cal L}/de{\omega_q^{\;\;\mu\nu}}\,ds +
\de{\cal L}/de{T_{qk}^\mu}e^\sigma_k\eta_{\sigma\nu}\,ds
+ \frac{1}{2}d\left(\de{\cal L}/de{R_{qk}^{\;\;\;\;\mu\nu}}\right)\wedge ds_k + \de{\cal L}/de{R_{kq}^{\;\;\;\;\alpha\nu}}\omega_{k\;\;\;\mu}^{\;\;\alpha}\,ds \right]G^{\mu\nu}_q \\
-{\cal J}\/(X)\interior d\left(\de{\cal L}/de{T_{hk}^\alpha}\right)\wedge\left(de^\alpha_h \wedge ds_k - \omega_{h\;\;\;\nu}^{\;\;\alpha}e^\nu_k\,ds + \frac{1}{2}T_{hk}^\alpha\,ds\right)\\
-\frac{1}{2}{\cal J}\/(X)\interior d\left(\de{\cal
L}/de{R_{hk}^{\;\;\;\;\alpha\beta}}\right)\wedge\left(d\omega_h^{\;\;\alpha\beta}\wedge
ds_k -
\omega_{h\;\;\;\lambda}^{\;\;\alpha}\omega_k^{\;\;\lambda\beta}\,ds
+ \frac{1}{2}R_{hk}^{\;\;\;\;\alpha\beta}\,ds\right)
\end{split}
\end{equation}
Due to the arbitrariness  of $X\/$ and the holonomy of the $\cal
J$-extension ${\cal J}\sigma\/$ (compare with eqs. (\ref{0.2}b)),
the requirement \eqref{1.4} yields two sets of final field
equations
\begin{subequations}\label{1.7}
\begin{equation}\label{1.7a}
{\cal J}\sigma^*\left(\de{\cal L}/de{e^\mu_q} + \de{\cal
L}/de{T_{kq}^\alpha}\omega_{k\;\;\;\mu}^{\;\;\alpha} \right) -\de
/de{x^k}\left({\cal J}\sigma^*\left(\de{\cal
L}/de{T_{kq}^\mu}\right)\right) =0
\end{equation}
and
\begin{equation}\label{1.7b}
\begin{split}
{\cal J}\sigma^*\left( \de{\cal L}/de{\omega_q^{\;\;\mu\nu}} - \de{\cal L}/de{T_{kq}^\mu}e^\sigma_k\eta_{\sigma\nu} + \de{\cal L}/de{T_{kq}^\nu}e^\sigma_k\eta_{\sigma\mu} + \de{\cal L}/de{R_{kq}^{\;\;\;\;\alpha\nu}}\omega_{k\;\;\;\mu}^{\;\;\alpha} + \de{\cal L}/de{R_{kq}^{\;\;\;\;\mu\alpha}}\omega_{k\;\;\;\nu}^{\;\;\alpha}\right)\\
-\de /de{x^k}\left({\cal J}\sigma^*\left(\de{\cal
L}/de{R_{kq}^{\;\;\;\;\mu\nu}} \right)\right) =0
\end{split}
\end{equation}
\end{subequations}
To conclude, it is worth  noticing that all the restrictions about
the vector fields ${\cal J}\/(X)\/$ in eq. \eqref{1.4} may be
removed. In fact, it is easily seen that eq. \eqref{1.4}
automatically implies
\begin{equation}\label{1.14}
{\cal J}\sigma^*\/(X\interior d\Theta)=0, \quad\quad \forall X\in
D^1\/({\cal J}\/(\E\times\C))
\end{equation}

\section{$f(R)$-gravity within the $\cal J$-bundle framework}

\subsection{Field equations in empty space}
Let us now apply the above formalism to the $f\/(R)$ theories of
gravity. The Lagrangian densities which we are going to consider
are of the specific kind ${\cal L}=ef\/(R)\/$, with
$e=\det(e^\mu_i)\/$ and
$R=R_{ij}^{\;\;\;\;\mu\nu}e^i_{\mu}e^j_{\nu}\/$. Therefore, taking
the identities ${\displaystyle \de e/de{e^\mu_i}=ee^i_\mu\/}$ and
${\displaystyle \de{e^j_\nu}/de{e^\mu_i}=-e^i_{\nu}e^j_{\mu}\/}$
into account, we have
\begin{subequations}
\begin{equation}\label{1.8a}
\de{\cal L}/de{e^\mu_i}=ee^i_{\mu}f\/(R) -
2ef'\/(R)R_{\mu\sigma}^{\;\;\;\;\lambda\sigma}e^{i}_\lambda
\end{equation}
\begin{equation}\label{1.8b}
\de{\cal
L}/de{R_{ki}^{\;\;\;\;\mu\nu}}=2ef'\/(R)\left[e^k_{\mu}e^i_{\nu} -
e^i_{\mu}e^k_{\nu}\right]
\end{equation}
\end{subequations}
In view of this, eqs. \eqref{1.7} become
\begin{subequations}\label{1.9}
\begin{equation}\label{1.9a}
e^i_{\mu}f\/(R) -
2f'\/(R)R_{\mu\sigma}^{\;\;\;\;\lambda\sigma}e^{i}_\lambda =0
\end{equation}
and
\begin{equation}\label{1.9b}
\begin{split}
\de/de{x^k}\left[2ef'\/(R)\left(e^k_{\mu}e^i_{\nu} - e^i_{\mu}e^k_{\nu}\right)\right] - \omega_{k\;\;\;\mu}^{\;\;\lambda}\left[2ef'\/(R)\left(e^k_{\lambda}e^i_{\nu} - e^i_{\lambda}e^k_{\nu}\right)\right]\\
-
\omega_{k\;\;\;\nu}^{\;\;\lambda}\left[2ef'\/(R)\left(e^k_{\mu}e^i_{\lambda}
- e^i_{\mu}e^k_{\lambda}\right)\right] =0
\end{split}
\end{equation}
\end{subequations}
After some calculations, eqs. \eqref{1.9b} may be rewritten in the
form
\begin{equation}\label{1.10}
ef''\/(R)\de{R}/de{x^t}e^\alpha_s -
ef''\/(R)\de{R}/de{x^s}e^\alpha_t -ef'\/(R)\left( T^\alpha_{ts} -
T^\sigma_{t\sigma}e^\alpha_s + T^\sigma_{s\sigma}e^\alpha_t
\right) =0
\end{equation}
where ${\displaystyle T^\alpha_{ts} = \de{e^\alpha_s}/de{x^t} -
\de{e^\alpha_t}/de{x^s} +
\omega^{\;\;\alpha}_{t\;\;\;\lambda}e^\lambda_s -
\omega^{\;\;\alpha}_{s\;\;\;\lambda}e^\lambda_t\/}$ are the
torsion coefficients of the connection
$\omega_i^{\;\;\mu\nu}\/(x)\/$.

Recalling the relationships
$R^h_{\;\;kij}=R_{ij}^{\;\;\;\;\mu\sigma}\eta_{\sigma\nu}e^h_\mu
e^\nu_k\/$ and $T_{ij}^{\;\;\;h}=T^\mu_{ij}e_\mu^h\/$ among the
quantities related to the spin connection $\omega\/$  and the
associated linear connection $\Gamma\/$, that is ${\displaystyle
\Gamma_{ij}^{\;\;\;h}=e^h_\mu\left(\de{e^\mu_j}/de{x^i} +
\omega_{i\;\;\;\nu}^{\;\;\mu}e^\nu_j\right)\/}$, it is
straightforward  to see that eqs. \eqref{1.9a} and \eqref{1.10}
are equivalent to eqs. \eqref{00.5} obtained in the metric--affine
formalism.

At this point, the  same considerations made in \cite{CCSV1} hold.
In particular, let us take into account the trace of the equation
\eqref{1.9a}, namely
\begin{equation}\label{1.11}
2f\/(R) - f'\/(R)R =0
\end{equation}
This is  identically satisfied by all possible values of $R\/$
only in the special case $f\/(R)=kR^2\/$. In all the other cases,
equation \eqref{1.11} represents a constraint on the scalar
curvature $R\/$. As a conclusion, it follows that, if $f\/(R)\not
= kR^2\/$, the scalar curvature $R\/$ has to be a constant (at
least on connected domains) and coincides with a given solution
value of \eqref{1.11}. In such a circumstance, equations
\eqref{1.10} imply that the torsion $T^\alpha_{ij}\/$ has to be
zero and the theory reduces to a $f\/(R)$-theory without torsion,
thus leading to Einstein equations with a cosmological constant.

In particular, we it is worth noticing that:

\bigskip\noindent
$\bullet\/$ in the case $f\/(R)=R\/$, eq. \eqref{1.11} yields
$R=0\/$  and therefore eqs. \eqref{1.9a} are equivalent to
Einstein's equations in empty space;

\bigskip\noindent
$\bullet\/$ if we assume $f\/(R)=kR^2\/$, by replacing eq.
\eqref{1.11} into eq. \eqref{1.9},  we obtain final field
equations of the form
\begin{subequations}\label{1.12}
\begin{equation}\label{1.12a}
\frac{1}{4}e^i_{\mu}R - R_\mu^{\;\;\lambda}e^i_\lambda =0
\end{equation}
\begin{equation}\label{1.12b}
\frac{1}{R}\de{R}/de{x^t}e^\alpha_s -
\frac{1}{R}\de{R}/de{x^s}e^\alpha_t  -\left( T^\alpha_{ts} -
T^\sigma_{t\sigma}e^\alpha_s + T^\sigma_{s\sigma}e^\alpha_t
\right) =0
\end{equation}
\end{subequations}
After some straightforward calculations,  eq. \eqref{1.12b} can be
put in normal form with respect to the torsion, namely
\begin{equation}\label{1.13}
T^\alpha_{ts}= -\frac{1}{2R}\de{R}/de{x^t}e^\alpha_s +
\frac{1}{2R}\de{R}/de{x^s}e^\alpha_t
\end{equation}

\subsection{Symmetries and conserved quantities}
The Poincar\'e--Cartan formulation \eqref{1.14} of the field
equations turns out to be especially useful in the study of
symmetries and conserved quantities. To see this point, we recall
the following  \cite{VC}
\begin{Definition}\label{Def2.1}
A vector field $Z\/$ on ${\cal J}\/(\E\times\C)\/$ is  called a
generalized infinitesimal Lagrangian symmetry if it satisfies the
requirement
\begin{equation}\label{2.1}
L_Z\/({\cal L}\,ds)=d\alpha
\end{equation}
for some $3$-form $\alpha\/$ on ${\cal J}\/(\E\times\C)\/$.
\end{Definition}
\begin{Definition}\label{Def2.2}
A vector field $Z\/$ on ${\cal J}\/(\E\times\C)\/$ is  called a
Noether vector field if it satisfies the condition
\begin{equation}\label{2.3}
L_Z\Theta = \omega + d\alpha
\end{equation}
where $\omega\/$ is a $4$-form belonging to the  ideal generated
by the contact forms and $\alpha\/$ is any $3$-form on ${\cal
J}\/(\E\times\C)\/$.
\end{Definition}
\begin{Proposition}\label{Pro2.1}
If a generalized infinitesimal Lagrangian symmetry  $Z\/$ is a
$\cal J$-prolongation, then it is a Noether vector field.
\end{Proposition}
\begin{Proposition}\label{Pro2.2}
If a Noether vector field $Z\/$ is a $\cal J$-prolongation , then
it is an infinitesimal dynamical symmetry.
\end{Proposition}
We can associate with any Noether vector field $Z\/$  a
corresponding conserved current. In fact, given $Z\/$ satisfying
eq. \eqref{2.3} and a critical section $\sigma : {\cal M}\to
\E\times\C\/$ we have
\begin{equation}\label{2.4}
d{\cal J}\sigma^*\/(Z\interior\Theta - \alpha) = {\cal
J}\sigma^*\/(\omega - Z\interior d\Theta)=0
\end{equation}
showing that the current ${\cal J}\sigma^*\/(Z\interior\Theta -
\alpha)\/$ is conserved on shell.

As it is well known, diffeomorphisms and Lorentz  transformations
(for tetrad and connection) have to be dynamical symmetries for
the theory: let us prove it.

To start with, let ${\displaystyle Y=\xi^i\,\de/de{x^i}\/}$ be the
generator of a (local) one parameter group of diffeomorphisms on
${\cal M}\/$. The vector field $Y\/$ may be ``lifted'' to a vector
field $X\/$ on $\E\times\C\/$ by setting
\begin{equation}\label{2.5}
X=\xi^i\,\de/de{x^i} - \de{\xi^k}/de{x^q}e^\mu_k\,\de/de{e^\mu_k}
-
\frac{1}{2}\de{\xi^k}/de{x^q}\omega_k^{\;\;\mu\nu}\,\de/de{\omega_q^{\;\;\mu\nu}}
\end{equation}
The vector fields \eqref{2.5} are $\cal J$-prolongable and their
$\cal J$-prolongations are expressed as \cite{VC}
\begin{equation}\label{2.6}
{\cal J}\/(X) = \xi^i\,\de/de{x^i} -
\de{\xi^k}/de{x^q}e^\mu_k\,\de/de{e^\mu_k} -
\frac{1}{2}\de{\xi^k}/de{x^q}\omega_k^{\;\;\mu\nu}\,\de/de{\omega_q^{\;\;\mu\nu}}
+ T^\mu_{jk}\de{\xi^k}/de{x^i}\,\de/de{T^\mu_{ij}} +
\frac{1}{2}R_{jk}^{\;\;\;\;\mu\nu}\de{\xi^k}/de{x^i}\,\de/de{R_{ij}^{\;\;\;\;\mu\nu}}
\end{equation}
A direct calculation shows that the vector fields \eqref{2.6}
satisfy $L_{{\cal J}\/(X)}\/(ef\/(R)\,ds)=0\/$, so proving that
they are infinitesimal Lagrangian symmetries for generic $f(R)$-
models. Due to Propositions \eqref{Pro2.1} and \eqref{Pro2.2}, we
conclude that the vector fields \eqref{2.6} are Noether vector
fields and thus infinitesimal dynamical symmetries. There are no
associated conserved quantities, the inner product ${\cal
J}\/(X)\interior\Theta\/$ consisting in an exact term plus a term
vanishing identically when pulled--back under critical section. We
have indeed
\begin{equation}\label{2.6bis}
\begin{split}
{\cal J}\/(X)\interior\Theta = e\xi^j\/\left(f\/(R)\delta^k_j - 2f'\/(R)R^k_j\right)ds_k - \frac{1}{4}\theta^{\alpha\beta}\wedge\left({\cal J}\/(X)\interior\de{\cal L}/de{R_{hk}^{\;\;\;\;\alpha\beta}}ds_{hk}\right)+ \\
\frac{1}{4}d\left(\xi^j\omega_j^{\;\;\alpha\beta}\de{\cal
L}/de{R_{hk}^{\;\;\;\;\alpha\beta}}ds_{hk}\right) -
\frac{1}{4}\xi^j\omega_j^{\;\;\alpha\beta}D\left(\de{\cal
L}/de{R_{hk}^{\;\;\;\;\alpha\beta}}\right)\wedge ds_{hk}
\end{split}
\end{equation}
where ${\displaystyle D\left(\de{\cal
L}/de{R_{hk}^{\;\;\;\;\alpha\beta}}\right)= d\left(\de{\cal
L}/de{R_{hk}^{\;\;\;\;\alpha\beta}}\right) - \de{\cal
L}/de{R_{hk}^{\;\;\;\;\lambda\beta}}\omega_{i\;\;\;\alpha}^{\;\;\lambda}\,dx^i
- \de{\cal
L}/de{R_{hk}^{\;\;\;\;\alpha\lambda}}\omega_{i\;\;\;\beta}^{\;\;\lambda}\,dx^i\/}$.

Infinitesimal Lorentz transformations are represented by  vector
fields on ${\cal J}\/(\E\times\C)\/$ of the form
\begin{equation}\label{2.7}
Y=A^\gamma_{\;\;\sigma}e^\sigma_q\,\de/de{e^\gamma_q} -
\frac{1}{2}D_qA^{\mu\nu}\,\de/de{\omega_q^{\;\;\mu\nu}}
\end{equation}
where $A^{\mu\nu}\/(x)=-A^{\nu\mu}\/(x)\/$ is a tensor--valued
function on ${\cal M}$ in the Lie algebra of $SO(3,1)\/$, and
${\displaystyle D_qA^{\mu\nu}=\de A^{\mu\nu}/de{x^q} +
\omega_{q\;\;\;\sigma}^{\;\;\mu}A^{\sigma\nu} +
\omega_{q\;\;\;\sigma}^{\;\;\nu}A^{\mu\sigma}\/}$. As above, the
vector fields \eqref{2.7} are $\cal J$-prolongable and their $\cal
J$-prolongations are expressed as  \cite{VC}
\begin{equation}\label{2.8}
{\cal J}\/(Y)=A^\gamma_{\;\;\sigma}e^\sigma_q\,\de/de{e^\gamma_q}
- \frac{1}{2}D_qA^{\mu\nu}\,\de/de{\omega_q^{\;\;\mu\nu}} +
\frac{1}{2}A^\mu_{\;\;\sigma}T^\sigma_{ij}\,\de/de{T^\mu_{ij}} +
\frac{1}{2}A^\mu_{\;\;\sigma}R_{ij}^{\;\;\;\;\sigma\nu}\,\de/de{R_{ij}^{\;\;\;\;\mu\nu}}
\end{equation}
It is  straightforward  to verify that the vector fields
\eqref{2.8} obey the condition $L_{{\cal
J}\/(Y)}\/(ef\/(R)\,ds)=0\/$. Once again, the conclusion follows
that they are infinitesimal dynamical symmetries. Moreover, it is
esily seen that
\begin{equation}\label{2.9}
{\cal J}\/(Y)\interior\Theta =
-\frac{1}{4}d\left(A^{\alpha\beta}\de{\cal
L}/de{R_{hk}^{\;\;\;\;\alpha\beta}}\,ds_{hk}\right) +
\frac{1}{4}A^{\alpha\beta}D\left(\de{\cal
L}/de{R_{hk}^{\;\;\;\;\alpha\beta}}\right)\wedge ds_{hk}
\end{equation}
Therefore, as above, since the  inner product ${\cal
J}\/(X)\interior\Theta\/$ consists in an exact term plus a term
vanishing identically when pulled--back under critical section,
there are no associated conserved quantities.

\subsection{Field equations in presence of matter}
In presence of  matter,  the configuration space--time of the
theory results to be the fiber product $\E\times\C\times_{\cal M}
F\/$ over ${\cal M}$, between $\E\times\C\/$ and the bundle $F\to
{\cal M}$ where the matter fields $\psi^A\/$ take their values.

The field equations  are derived from a variational problem built
on the manifold $\J\times_{\cal M} J_1\/(F)\/$, where $J_1\/(F)\/$
indicates the standard first ${\cal J}$-bundle associated with the
fibration $F\to {\cal M}$.

The total Lagrangian  density of the theory is obtained by adding
to the gravitational one  a suitable matter Lagrangian density
${\cal L}_m\/$. Throughout the paper, we shall consider matter
Lagrangian densities of the kind ${\cal L}_m={\cal
L}_m\/(e,\omega,\psi,\partial\psi)\/$. The corresponding
Poincar\'e--Cartan form is given by the sum $\Theta + \theta_m\/$,
where ${\displaystyle \theta_m={\cal L}_m\,ds + \de{{\cal
L}_m}/de{\psi^A_i}\,\theta^A\wedge ds_i}$ is the standard
Poincar\'e--Cartan form associated with the matter density ${\cal
L}_m\/$, being $\theta^A=d\psi^A - \psi^A_i\,ds_i\/$  the usual
contact $1$-forms of the bundle $J_1\/(F)\/$.

In such a circumstance, the Euler--Lagrange  equations \eqref{1.7}
assume the local expression
\begin{subequations}\label{2.2.2}
\begin{equation}\label{2.2.2a}
f'\/(R)R_{\mu\sigma}^{\;\;\;\;\lambda\sigma}e^{i}_\lambda
-\frac{1}{2}e^i_{\mu}f\/(R)=\Sigma^i_{\;\mu}
\end{equation}
and
\begin{equation}\label{2.2.2b}
f'\/(R)\left( T^\alpha_{ts} - T^\sigma_{t\sigma}e^\alpha_s +
T^\sigma_{s\sigma}e^\alpha_t \right) =\de{f'(R)}/de{x^t}e^\alpha_s
- \de{f'(R)}/de{x^s}e^\alpha_t + S^\alpha_{ts}
\end{equation}
\end{subequations}
where ${\displaystyle \Sigma^i_{\;\mu} :=
\frac{1}{2e}\frac{\partial{\cal L}_m}{\partial e^\mu_i}\/}$ and
${\displaystyle S^\alpha_{ts}:=- \frac{1}{2e}\frac{\partial{\cal
L}_m}{\partial\omega_i^{\;\;\mu\nu}}e^\mu_t e^\nu_s e^\alpha_i\/}$
play the role of energy--momentum and spin density tensors
respectively. In particular, from eqs. \eqref{2.2.2b}, we obtain
\begin{equation}\label{2.2.3}
f'T_{t\sigma}^\sigma = -\frac{3}{2}\de{f'}/de{x^t} -
\frac{1}{2}S_{t\sigma}^{\sigma}
\end{equation}
Then, substituting eqs. \eqref{2.2.3} into eqs. \eqref{2.2.2b}, we
find the expression for the torsion
\begin{equation}\label{2.2.4}
T^\alpha_{ts}=-\frac{1}{2f'}\left(\de{f'}/de{x^p} +
S_{p\sigma}^\sigma\right)\left( \delta^p_{t}e^\alpha_s -
\delta^p_{s}e^\alpha_t \right) + \frac{1}{f'}S_{ts}^\alpha
\end{equation}
Eqs. \eqref{2.2.4} tell us that, in presence of $\omega$-dependent
matter, there are two sources of torsion: the spin density
$S_{ts}^\alpha\/$ and the nonlinearity of the gravitational
Lagrangian. It is important to stress that this feature is not
present in standard GR.

Now, by considering the  trace of eqs. \eqref{2.2.2a} we obtain a
relation between the scalar curvature $R\/$ and the trace
$\Sigma\/$ of the energy--momentum tensor given by
\begin{equation}\label{2.2.5}
f'\/(R)R -2f\/(R) = \Sigma
\end{equation}
When the trace $\Sigma\/$ is allowed to assume only a constant
value,  the present theory amounts to an Einstein--like (if
$S^\alpha_{ts}=0$, i.e. $\omega$-independent matter) or an
Einstein--Cartan--like theory (if $S^\alpha_{ts}\not= 0$, i.e.
$\omega$-dependent matter) with cosmological constant. In fact, in
such a circumstance, eq. \eqref{2.2.5} implies that the scalar
curvature $R\/$ also is constant. As a consequence, eqs.
\eqref{2.2.2a} and \eqref{2.2.4} can be expressed as
\begin{subequations}\label{2.2.6}
\begin{equation}\label{2.2.6a}
R_{\;\mu}^{i} - \frac{1}{2}\/\left(R + \Lambda \right)\/e^i_\mu =
k\Sigma_{\;\mu}^{i}
\end{equation}
\begin{equation}\label{2.2.6b}
T^\alpha_{ts}= \frac{k}{2}\/\left( 2S_{ts}^\alpha -
S_{t\sigma}^{\sigma}e^\alpha_s + S_{s\sigma}^{\sigma}e^\alpha_t
\right)
\end{equation}
\end{subequations}
where $\Lambda= kf\/(R) - R\/$ and ${\displaystyle
k=\frac{1}{f'\/(R)}\/}$, $R\/$ being the constant value determined
by eq.\eqref{2.2.5}, provided that $f'(R)\not= 0\/$. The previous
discussion holds with the exception of the particular case
$\Sigma=0\/$ and $f(R)=\alpha\/R^2\/$. Indeed, under these
conditions, eq. \eqref{2.2.5} is a trivial identity which imposes
no restriction on the scalar curvature $R\/$.

From now on,  we shall  suppose that $\Sigma\/$ is not forced to
be a constant  when the matter field equations are satisfied.
Besides, we shall suppose that the relation \eqref{2.2.5} is
invertible so that the scalar curvature can be thought as a
suitable function of $\Sigma\/$, namely
\begin{equation}\label{2.2.7}
R=F\/(\Sigma)
\end{equation}
With this assumption in mind, defining the  tensors
$R^i_{\;j}:=R_{\mu\sigma}^{\;\;\;\;\lambda\sigma}e^{i}_{\lambda}e^\mu_j$,
$\Sigma^i_{\;j}:=\Sigma^i_{\;\mu}e^\mu_j$,
$T_{ij}^{\;\;\;h}:=T^\alpha_{ij}e_\alpha^h\/$ and
$S^{\;\;\;h}_{ij}:=S^\alpha_{ij}e_\alpha^h\/$, we rewrite eqs.
\eqref{2.2.2a} and \eqref{2.2.4} in the equivalent form
\begin{subequations}\label{2.2.8}
\begin{equation}\label{2.2.8a}
R_{ij} -\frac{1}{2}Rg_{ij}= \frac{1}{f'\/(F\/(\Sigma))}\left(
\Sigma_{ij} - \frac{1}{4}\Sigma g_{ij} \right) -
\frac{1}{4}F\/(\Sigma)g_{ij}
\end{equation}
\begin{equation}\label{2.2.8b}
T_{ij}^{\;\;\;h} = -
\frac{1}{2f'\/(F\/(\Sigma))}\left(\de{f'\/(F\/(\Sigma))}/de{x^p} +
S_{p\sigma}^\sigma\right)\left(\delta^p_i\delta^h_j -
\delta^p_j\delta^h_i\right) +
\frac{1}{f'\/(F\/(\Sigma))}S^{\;\;\;h}_{ij}
\end{equation}
\end{subequations}
In eqs. \eqref{2.2.8a} one has to distinguish the order of the
indexes since, in general, the tensors $R_{ij}\/$ and
$\Sigma_{ij}\/$ are not symmetric.

Moreover, following \cite{CCSV1}, in the l.h.s. of eqs.
\eqref{2.2.8a}, we can distinguish the contribution due to the
Christoffel terms from that due to the torsion dependent terms. To
see this point, from eqs. \eqref{00.2} and \eqref{00.4}, we first
get the following representation for the contracted curvature
tensor
\begin{equation}\label{2.2.9}
R_{ij}=\tilde{R}_{ij} + \tilde{\nabla}_jK_{hi}^{\;\;\;h} -
\tilde{\nabla}_hK_{ji}^{\;\;\;h} +
K_{ji}^{\;\;\;p}K_{hp}^{\;\;\;h} -
K_{hi}^{\;\;\;p}K_{jp}^{\;\;\;h}
\end{equation}
where $\tilde{R}_{ij}\/$ is the Ricci tensor of the Levi--Civita
connection $\tilde\Gamma\/$ associated with the metric
$g_{ij}=\eta_{\mu\nu}e^\mu_{i}e^\nu_{j}\/$, and $\tilde{\nabla}\/$
denotes the Levi--Civita covariant derivative. Then, recalling the
expression of the contortion tensor \cite{Hehl}
\begin{equation}\label{2.2.10}
K_{ij}^{\;\;\;h} = \frac{1}{2}\/\left( - T_{ij}^{\;\;\;h} +
T_{j\;\;\;i}^{\;\;h} - T^h_{\;\;ij}\right)
\end{equation}
and using the second set of field equations \eqref{2.2.8b}, we
obtain the following representations
\begin{subequations}\label{2.2.11}
\begin{equation}\label{2.2.11a}
K_{ij}^{\;\;\;h}= \hat{K}_{ij}^{\;\;\;h} + \hat{S}_{ij}^{\;\;\;h}
\end{equation}
\begin{equation}\label{2.2.11b}
\hat{S}_{ij}^{\;\;\;h}:=\frac{1}{2f'}\/\left( - S_{ij}^{\;\;\;h} +
S_{j\;\;\;i}^{\;\;h} - S^h_{\;\;ij}\right)
\end{equation}
\begin{equation}\label{2.2.11c}
\hat{K}_{ij}^{\;\;\;h} := -\hat{T}_j\delta^h_i +
\hat{T}_pg^{ph}g_{ij}
\end{equation}
\begin{equation}\label{2.2.11d}
\hat{T}_j:=\frac{1}{2f'}\/\left( \de{f'}/de{x^j} +
S^\sigma_{j\sigma} \right)
\end{equation}
\end{subequations}
Inserting eqs. \eqref{2.2.11} in eq. \eqref{2.2.9} we end up with
the final expression for $R_{ij}\/$
\begin{equation}\label{2.2.12}
\begin{split}
R_{ij}=\tilde{R}_{ij} + \tilde{\nabla}_j\hat{K}_{hi}^{\;\;\;h} +
\tilde{\nabla}_j\hat{S}_{hi}^{\;\;\;h} -
\tilde{\nabla}_h\hat{K}_{ji}^{\;\;\;h} - \tilde{\nabla}_h\hat{S}_{ji}^{\;\;\;h} + \hat{K}_{ji}^{\;\;\;p}\hat{K}_{hp}^{\;\;\;h} + \hat{K}_{ji}^{\;\;\;p}\hat{S}_{hp}^{\;\;\;h} \\
+ \hat{S}_{ji}^{\;\;\;p}\hat{K}_{hp}^{\;\;\;h} +
\hat{S}_{ji}^{\;\;\;p}\hat{S}_{hp}^{\;\;\;h} -
\hat{K}_{hi}^{\;\;\;p}\hat{K}_{jp}^{\;\;\;h} -
\hat{K}_{hi}^{\;\;\;p}\hat{S}_{jp}^{\;\;\;h} -
\hat{S}_{hi}^{\;\;\;p}\hat{K}_{jp}^{\;\;\;h} -
\hat{S}_{hi}^{\;\;\;p}\hat{S}_{jp}^{\;\;\;h}
\end{split}
\end{equation}
The last step is the substitution of eqs. \eqref{2.2.12} into eqs.
\eqref{2.2.8a}. Explicit examples of the  described procedure are
given in the next subsections. We will discuss specific cases of
fields  coupled with gravity acting as matter sources.

\subsection{The case of Dirac fields}
As a first example of matter, we consider  the case of Dirac
fields $\psi\/$. The matter Lagrangian density is given by
\begin{equation}\label{2.4.1}
{\cal L}_D = e\/\left[ \frac{i}{2}\/\left( \bar\psi\gamma^iD_i\psi
- D_i\bar\psi\gamma^i\psi \right) - m\bar\psi\psi \right]
\end{equation}
where ${\displaystyle D_i\psi = \de\psi/de{x^i} +
\omega_i^{\;\;\mu\nu}S_{\mu\nu}\psi\/}$ and ${\displaystyle
D_i\bar\psi = \de{\bar\psi}/de{x^i} -
\bar\psi\omega_i^{\;\;\mu\nu}S_{\mu\nu}\/}$ are the covariant
derivatives of the Dirac fields, ${\displaystyle
S_{\mu\nu}=\frac{1}{8}\left[\gamma_\mu,\gamma_\nu\right]\/}$,
$\gamma^i =\gamma^{\mu}e^i_\mu\/$; $\gamma^\mu\/$ denotes the
Dirac matrices.

The field equations for the Dirac fields are
\begin{equation}\label{2.4.1bis}
i\gamma^hD_h\psi - m\psi =0, \qquad iD_h\bar{\psi}\gamma^h +
m\bar\psi =0
\end{equation}
Since the Lagrangian \eqref{2.4.1} vanishes  for $\psi\/$ and
$\bar\psi\/$ satisfying the equation \eqref{2.4.1bis}, the
corresponding energy--momentum and spin density tensors are
expressed, respectively, as
\begin{equation}\label{2.4.2}
\Sigma_{ij} = \frac{i}{4}\/\left[ \bar\psi\gamma_iD_j\psi -
\left(D_j\bar\psi\right)\gamma_i\psi \right]
\end{equation}
and
\begin{equation}\label{2.4.3}
S_{ij}^{\;\;\;h}=
-\frac{i}{2}\bar\psi\left\{\gamma^h,S_{ij}\right\}\psi
\end{equation}
with ${\displaystyle
S_{ij}=\frac{1}{8}\left[\gamma_i,\gamma_j\right]\/}$. Now, using
the properties of the Dirac matrices, it is easily seen that
${\displaystyle
\left\{\gamma^h,S^{ij}\right\}=\frac{1}{2}\gamma^{[i}\gamma^j\gamma^{h]}\/}$.
This fact implies the total antisymmetry of the spin density
tensor $S_{ij}^{\;\;\;h}\/$. As a consequence, the contracted
curvature and  scalar curvature assume the simplified expressions
(compare with eq. \eqref{2.2.12})
\begin{subequations}\label{2.4.4}
\begin{equation}\label{2.4.4a}
R_{ij} = \tilde{R}_{ij} - 2\tilde{\nabla}_{j}\hat{T}_i -
\tilde{\nabla}_h\hat{T}^hg_{ij} + 2\hat{T}_i\hat{T}_j -
2\hat{T}_h\hat{T}^hg_{ij} - \tilde{\nabla}_h\hat{S}_{ji}^{\;\;\;h}
- \hat{S}_{hi}^{\;\;\;p}\hat{S}_{jp}^{\;\;\;h}
\end{equation}
and
\begin{equation}\label{2.4.4b}
R = \tilde{R} - 6\tilde{\nabla}_{i}\hat{T}^i - 6\hat{T}_i\hat{T}^i
- \hat{S}_{hi}^{\;\;\;p}\hat{S}_{\;\;p}^{i\;\;\;h}
\end{equation}
\end{subequations}
where now ${\displaystyle \hat{T}_i
=\frac{1}{2f'}\de{f'}/de{x^i}\/}$ and ${\displaystyle
\hat{S}_{ij}^{\;\;\;h}:=-\frac{1}{2f'}S_{ij}^{\;\;\;h}\/}$.
Inserting eqs. \eqref{2.4.4} into eqs. \eqref{2.2.8a} and  using
of the above expression for $\hat{T}_i\/$, we obtain the final
Einstein--like equations
\begin{equation}\label{2.4.5}
\begin{split}
\tilde{R}_{ij} -\frac{1}{2}\tilde{R}g_{ij}= \frac{1}{\varphi}\Sigma_{ij} + \frac{1}{\varphi^2}\left( - \frac{3}{2}\de\varphi/de{x^i}\de\varphi/de{x^j} + \varphi\tilde{\nabla}_{j}\de\varphi/de{x^i} + \frac{3}{4}\de\varphi/de{x^h}\de\varphi/de{x^k}g^{hk}g_{ij} \right. \\
\left. - \varphi\tilde{\nabla}^h\de\varphi/de{x^h}g_{ij} -
V\/(\varphi)g_{ij} \right) +
\tilde{\nabla}_h\hat{S}_{ji}^{\;\;\;h} +
\hat{S}_{hi}^{\;\;\;p}\hat{S}_{jp}^{\;\;\;h} -
\frac{1}{2}\hat{S}_{hq}^{\;\;\;p}\hat{S}_{\;\;p}^{q\;\;\;h}g_{ij}
\end{split}
\end{equation}
where we have defined the scalar field
\begin{equation}\label{2.4.6}
\varphi := f'\/(F\/(\Sigma))
\end{equation}
and the effective potential
\begin{equation}\label{2.4.7}
V\/(\varphi):= \frac{1}{4}\left[ \varphi
F^{-1}\/((f')^{-1}\/(\varphi)) +
\varphi^2\/(f')^{-1}\/(\varphi)\right]
\end{equation}
To conclude, we notice that eqs. \eqref{2.4.5} can be simplified
by performing a conformal transformation. Indeed, setting
$\bar{g}_{ij}:=\varphi\eta_{\mu\nu}e^\mu_ie^\nu_j\/$, eqs.
\eqref{2.4.5} can be rewritten in the easier form
\begin{equation}\label{2.4.8}
\bar{R}_{ij} - \frac{1}{2}\bar{R}\bar{g}_{ij} =
\frac{1}{\varphi}\Sigma_{ij} -
\frac{1}{\varphi^3}V\/(\varphi)\bar{g}_{ij} +
\tilde{\nabla}_h\hat{S}_{ji}^{\;\;\;h} +
\hat{S}_{hi}^{\;\;\;p}\hat{S}_{jp}^{\;\;\;h} -
\frac{1}{2\varphi}\hat{S}_{hq}^{\;\;\;p}\hat{S}_{\;\;p}^{q\;\;\;h}g_{ij}
\end{equation}
where $\bar{R}_{ij}\/$  and $\bar{R}\/$ are respectively the Ricci
tensor and the Ricci scalar curvature associated with the
conformal metric $\bar{g}_{ij}\/$.

\subsection{The case of Yang--Mills fields}
As it has been shown in some previous works
\cite{CVB1,CVB2,VC,VCB2}, also gauge theories can be formulated
within the framework of ${\cal J}$-bundles. Therefore, we can
describe $f(R)$-gravity coupled with Yang--Mills fields  in the
new geometric setting.

To see this point,  let $Q\to {\cal M}$ be a principal fiber
bundle over space--time, with structural group a semisimple Lie
group $G\/$. We consider the affine bundle $J_1\/Q/G\to {\cal
M}\/$ (the space of principal connections of $Q\to {\cal M}\/$)
and refer it to local coordinates $x^i,a^A_i\/$,
$A=1,\ldots,r=\dim G\/$.

In a combined theory of $f\/(R)$-gravity and Yang--Mills fields,
additional dynamical fields are principal connections of $Q\/$
represented by sections of the bundle $J_1\/Q/G\to {\cal M}\/$.
The extended configuration space of the theory is then the fiber
product $\E\times\C\times J_1\/Q/G\to {\cal M}\/$ over ${\cal M}$.

Following the approach illustrated in Sec. II, we may construct
the quotient space ${\cal J}\/(\E\times\C\times J_1\/Q/G)\/$  (see
\cite{VC} for details). As additional $\cal J$-coordinates the
latter admits the components $F^A_{ij}\/$ of the curvature tensors
of the principal connections of $Q\to {\cal M}\/$. Holonomic
sections of the bundle ${\cal J}\/(\E\times\C\times J_1\/Q/G)\/$
are of the form \eqref{0.2} together with
\begin{equation}\label{3.1}
F^A_{ij}\/(x)=\de{a^A_j\/(x)}/de{x^i} - \de{a^A_i\/(x)}/de{x^j} +
a^B_j\/(x)a^C_i\/(x)C^A_{CB}
\end{equation}
where $C^A_{CB}\/$ are the structure coefficients of the Lie
algebra of $G\/$.

The Poincar\'e--Cartan $4$-form associated with  a Lagrangian on
${\cal J}\/(\E\times\C\times J_1\/Q/G)\/$ of the form $L={\cal
L}\/(x^i,e^\mu_i,a^A_i,R_{ij}^{\;\;\;\;\mu\nu},F^A_{ij})\,ds\/$ is
\begin{equation}\label{3.2}
\Theta = {\cal L}\, ds - \frac{1}{4}\de{\cal
L}/de{R_{hk}^{\;\;\;\;\alpha\beta}}\,\theta^{\alpha\beta}\wedge
ds_{hk} - \frac{1}{2}\de{\cal L}/de{F_{hk}^{A}}\,\theta^A\wedge
ds_{hk}
\end{equation}
where $\theta^A = \Phi^A - F^A\/$,  being $F^A :=
\frac{1}{2}F^A_{ij}\,dx^i\wedge dx^j\/$ and $\Phi^A :=
da^A_i\wedge dx^i + \frac{1}{2}a^B_ia^C_jC^A_{CB}\,dx^j\wedge
dx^i\/$.

Variational field equations are still of the form
\begin{equation}\label{3.3}
{\cal J}\sigma^*\left({\cal J}\/(X)\interior d\Theta\right)=0
\end{equation}
for all $\cal J$-prolongable vector  fields X on $\E\times\C\times
J_1\/Q/G\/$. Due to the arbitrariness of the infinitesimal
deformations $X\/$, eq. \eqref{3.3} splits into three sets of
final equations, respectively given by eqs. \eqref{1.7} together
with
\begin{equation}\label{3.4}
{\cal J}\sigma^*\left(\de{\cal L}/de{a^A_i} - D_k\de{\cal
L}/de{F^A_{kj}}\right)=0
\end{equation}
In particular, if the Lagrangian density  is
\begin{equation}
{\cal L}=e\/\left(f\/(R)
-\frac{1}{4}F^A_{ij}F^B_{pq}\gamma_{AB}\eta^{\mu\nu}e^p_{\mu}e^i_{\nu}\eta^{\lambda\sigma}e^q_{\lambda}e^j_{\sigma}
\right)\/\end{equation} expressing $f\/(R)\/$-gravity coupled with
a free Yang--Mills field, eqs. \eqref{1.7} and \eqref{3.4} assume
the explicit form
\begin{subequations}\label{3.5}
\begin{equation}\label{3.5a}
e^i_{\mu}f\/(R) -
2f'\/(R)R_{\mu\sigma}^{\;\;\;\;\lambda\sigma}e^{i}_\lambda
=\frac{1}{4}F^A_{jk}F_A^{jk}e^i_\mu - F^{Ai}_jF^j_{Ak}e^k_\mu
\end{equation}
\begin{equation}\label{3.5b}
\begin{split}
\de/de{x^k}\left[2ef'\/(R)\left(e^k_{\mu}e^i_{\nu} - e^i_{\mu}e^k_{\nu}\right)\right] - \omega_{k\;\;\;\mu}^{\;\;\lambda}\left[2ef'\/(R)\left(e^k_{\lambda}e^i_{\nu} - e^i_{\lambda}e^k_{\nu}\right)\right]\\
-
\omega_{k\;\;\;\nu}^{\;\;\lambda}\left[2ef'\/(R)\left(e^k_{\mu}e^i_{\lambda}
- e^i_{\mu}e^k_{\lambda}\right)\right] =0
\end{split}
\end{equation}
\begin{equation}\label{3.5c}
D_k\/(eF^{ik}_A)=0
\end{equation}
\end{subequations}
where ${\displaystyle D_k\/(eF^{ik}_A)=\de{(eF^{ik}_A)}/de{x^k}
-a^B_k\/(eF^{ik}_C)C^C_{BA}\/}$.

Since  the trace of the energy--impulse tensor ${\displaystyle
T^i_\mu := \frac{1}{4}F^A_{jk}F_A^{jk}e^i_\mu -
F^{Ai}_jF^j_{Ak}e^k_\mu\/}$ vanishes identically, we have again:

\bigskip\noindent
$\bullet\/$ if $f\/(R) =kR\/$ we recover the Einstein--Yang--Mills
theory;

\bigskip\noindent
$\bullet\/$ if $f\/(R) \not = kR^2\/$ the torsion is necessarily
zero and we recover a $f\/(R)$-theory without torsion coupled with
a Yang--Mills field;

\bigskip\noindent
$\bullet\/$ We can have non--vanishing torsion only in the case
$f\/(R)=kR^2\/$.

\subsection{The case of spin fluid matter}
As a last example of matter source, we consider the case of a
semiclassical spin fluid. This is characterized by an
energy--momentum tensor of the form
\begin{subequations}\label{2.5.1}
\begin{equation}\label{2.5.1a}
\Sigma^{ij}= (\rho + p)U^iU^j + pg^{ij}
\end{equation}
and a spin density tensor given by
\begin{equation}\label{2.5.1b}
S_{ij}^{\;\;\;h}=S_{ij}U^h
\end{equation}
\end{subequations}
where $U^i\/$  and $S_{ij}\/$ denote, respectively, the
$4$-velocity and the spin density of the fluid  (see, for example,
\cite{Hehl-Heyde-Kerlick} and references therein). However, the
constraint $U^iU_i=-1\/$ must hold. Other models of spin fluids
are possible, where, due to the treatment of spin as a
thermodynamical variable, different expressions for the
energy--momentum tensor may be taken into account
\cite{Ray-Smalley,Gasperini,deRitis}.

The $4$-velocity and the spin density satisfy by the so called
{\it convective condition}
\begin{equation}\label{2.5.2}
S_{ij}U^j =0
\end{equation}
It is easily seen that  the relations \eqref{2.5.2} imply the
identities
\begin{equation}\label{2.5.2bis}
\hat{S}_i^{\;\;ih}=-\hat{S}_i^{\;\;hi}
\end{equation}
obtained inserting eq. \eqref{2.5.1b} in eq. \eqref{2.2.11b} and
using \eqref{2.5.2}. Making use of \eqref{2.5.2bis} as well as of
eqs. \eqref{2.2.11b}, \eqref{2.2.12}, \eqref{2.5.1b} and
\eqref{2.5.2}, we can express the  Ricci curvature tensor and
scalar respectively as
\begin{subequations}\label{2.5.3}
\begin{equation}\label{2.5.3a}
\begin{split}
R_{ij} = \tilde{R}_{ij} - 2\tilde{\nabla}_{j}\hat{T}_i - \tilde{\nabla}_h\hat{T}^hg_{ij} + 2\hat{T}_i\hat{T}_j - 2\hat{T}_h\hat{T}^hg_{ij} - \frac{1}{f'}\hat{T}_hS^h_{\;\;j}U_i \\
- \frac{1}{2f'}\tilde{\nabla}_h\/\left(S_{ji}U^h + S_i^{\;\;h}U_j
- S^h_{\;\;j}U_i\right) + \frac{1}{4(f')^2}S^{pq}S_{pq}U_iU_j
\end{split}
\end{equation}
and
\begin{equation}\label{2.5.3b}
R = \tilde{R} - 6\tilde{\nabla}_{i}\hat{T}^i - 6\hat{T}_i\hat{T}^i
- \frac{1}{4(f')^2}S^{pq}S_{pq}
\end{equation}
\end{subequations}
where ${\displaystyle \hat{T}_i =\frac{1}{2f'}\de{f'}/de{x^i}\/}$.
In view of this, substituting eqs. \eqref{2.5.3} in eqs.
\eqref{2.2.8a},  and using the definitions \eqref{2.4.6} and
\eqref{2.4.7}, we obtain Einstein--like equations of the form
\begin{equation}\label{2.5.4}
\begin{split}
\tilde{R}_{ij} -\frac{1}{2}\tilde{R}g_{ij}= \frac{1}{\varphi}\Sigma_{ij} + \frac{1}{\varphi^2}\left( - \frac{3}{2}\de\varphi/de{x^i}\de\varphi/de{x^j} + \varphi\tilde{\nabla}_{j}\de\varphi/de{x^i} + \frac{3}{4}\de\varphi/de{x^h}\de\varphi/de{x^k}g^{hk}g_{ij} \right. \\
\left. - \varphi\tilde{\nabla}^h\de\varphi/de{x^h}g_{ij} - V\/(\varphi)g_{ij} \right) + \frac{1}{\varphi}\hat{T}_hS^h_{\;\;j}U_i + \frac{1}{2\varphi}\tilde{\nabla}_h\/\left(S_{ji}U^h + S_i^{\;\;h}U_j - S^h_{\;\;j}U_i\right) \\
- \frac{1}{4\varphi^2}S^{pq}S_{pq}U_iU_j -
\frac{1}{8\varphi^2}S^{pq}S_{pq}g_{ij}
\end{split}
\end{equation}
Eqs. \eqref{2.5.4} are the "microscopic" field equations for
$f(R)$ gravity with torsion, coupled  with a semiclassical spin
fluid. In this form, all the source contributions are put  in
evidence and their role is clearly defined into dynamics.

\subsection{Cosmological applications}

In order to apply the  above considerations to
Friedmann-Robertson-Walker(FRW) cosmological models, let us
consider an isotropic and homogeneous universe filled with a
cosmological spin fluid.

Eqs. \eqref{2.5.4} are valid in the microscopic domain of matter.
Cosmological equations can be derived by a suitable space--time
averaging of \eqref{2.5.4}. In this situation, the simplest
cosmological scenario is achieved by supposing that the
cosmological fluid is unpolarized. In fact, being the spin
randomly oriented, we can assume that the average of the spin and
its gradient vanish, but the same is not true for the
spin--squared terms  as $<S^{pq}S_{pq}>\/$. As conclusion, it
follows that, after averaging, eqs. \eqref{2.5.4} reduce to
\begin{equation}\label{2.5.5}
\begin{split}
\tilde{R}_{ij} -\frac{1}{2}\tilde{R}g_{ij}= \frac{1}{\varphi}\Sigma_{ij} + \frac{1}{\varphi^2}\left( - \frac{3}{2}\de\varphi/de{x^i}\de\varphi/de{x^j} + \varphi\tilde{\nabla}_{j}\de\varphi/de{x^i} + \frac{3}{4}\de\varphi/de{x^h}\de\varphi/de{x^k}g^{hk}g_{ij} \right. \\
\left. - \varphi\tilde{\nabla}^h\de\varphi/de{x^h}g_{ij} -
V\/(\varphi)g_{ij} \right) - \frac{1}{2\varphi^2}s^2\/U_iU_j -
\frac{1}{4\varphi^2}s^2g_{ij}
\end{split}
\end{equation}
where  we have introduced the notation $s^2 =2S^{pq}S_{pq}\/$ (see
\cite{Hehl-Heyde-Kerlick}).

As for the case of Dirac fields,  eqs. \eqref{2.5.5} can be
simplified by performing a conformal transformations. In view of
this,  we can  suppose that $\varphi>0\/$ where a sufficient
condition to satisfy this request is $f'>0\/$. Moreover, following
the line illustrated in \cite{CCSV1}, we introduce the vector
field ${\displaystyle \bar{U}^i:=\frac{U^i}{\sqrt{\varphi}}\/}$
representing the four velocity of the fluid with respect to the
conformal metric $\bar{g}_{ij}:=\varphi\/g_{ij}$ ($\bar{e}^\mu_i
:= \sqrt{\varphi}e^\mu_i\/$). The $1$-form $\bar{U}_i :=
\sqrt{\varphi}U_i\/$ denotes the corresponding covariant relation.

Then, performing the conformal transformation $\bar{e}^\mu_i :=
\sqrt{\varphi}e^\mu_i\/$, from \eqref{2.5.5} we obtain a set of
equivalent Einstein--like equations for the barred tetrad,
expressed as
\begin{equation}\label{2.5.6}
\bar{G}_{ij} = \frac{1}{\varphi^2}\bar{\Sigma}_{ij} -
\frac{1}{\varphi^3}V\/(\varphi)\bar{g}_{ij} -
\frac{1}{2\varphi^3}s^2\bar{U}_i\bar{U}_j -
\frac{1}{4\varphi^3}s^2\bar{g}_{ij}
\end{equation}
where $\bar{G}_{ij}\/$ is the Einstein tensor in the barred metric
and $\bar{\Sigma}_{ij}:=(\rho + p)\bar{U}_i\bar{U}_j +
p\bar{g}_{ij}\/$ the new stress-energy tensor for the perfect
fluid. Now, looking for a FRW solution
\begin{equation}\label{2.5.6bis}
\bar{e}^0=dt, \quad \bar{e}^1=a(t)\,d\psi, \quad
\bar{e}^2=a(t)\chi\,d\theta, \quad
\bar{e}^3=a(t)\chi\sin\theta\,d\phi
\end{equation}
for eqs. \eqref{2.5.6}, we get the Friedmann--like equations of
the form
\begin{subequations}\label{2.5.7}
\begin{equation}\label{2.5.7a}
3\left(\frac{\dot a}{a}\right)^2 + \frac{3k}{a^2} =
\frac{\rho}{\varphi^2} + \frac{V\/(\varphi)}{\varphi^3} -
\frac{1}{2\varphi^3}s^2 + \frac{1}{4\varphi^3}s^2
\end{equation}
and
\begin{equation}\label{2.5.7b}
-2\frac{\ddot{a}}{a} - \left(\frac{\dot a}{a}\right)^2 -
\frac{k}{a^2} = \frac{p}{\varphi^2} -
\frac{V\/(\varphi)}{\varphi^3} - \frac{1}{4\varphi^3}s^2
\end{equation}
\end{subequations}
In conclusion, we notice that, once a solution $\bar{e}^\mu\/$ is
found, also the conformal tetrad ${\displaystyle
e^\mu=\frac{1}{\sqrt{\varphi}}\bar{e}^\mu\/}$ (solution of
\eqref{2.5.5}) gives rise to a FRW metric. In fact, by performing
the time variable transformation
\begin{equation}\label{2.5.8}
d\tau := \frac{1}{\sqrt{\varphi\/(t)}}\,dt
\end{equation}
we can express the tetrad $e^\mu\/$ as
\begin{equation}\label{2.5.9}
e^0=d\tau, \quad e^1=A(\tau)\,d\psi, \quad
e^2=A(\tau)\chi\,d\theta, \quad e^3=A(\tau)\chi\sin\theta\,d\phi
\end{equation}
where ${\displaystyle A:=\frac{a}{\sqrt{\varphi}}\/}$.

\subsection{Equivalence with scalar--tensor theories}
The results of the last subsection lead to take into account the
analogies between $f(R)$-gravity with torsion and scalar-tensor
theories with torsion, as discussed, for example, in
\cite{German,Sung}. To this end, let us consider a Lagrangian
density of the form
\begin{equation}\label{2.6.1}
{\cal L}= \varphi\/eR -eU\/(\varphi) + {\cal L}_m
\end{equation}
where  $\varphi\/$ is a scalar field, $U\/(\varphi)$ is a suitable
potential and ${\cal L}_m\/$ is a  matter Lagrangian density.

The Euler--Lagrange equations \eqref{1.7} applied to the
Lagrangian density  \eqref{2.6.1}, yield the corresponding field
equations
\begin{subequations}\label{2.6.2}
\begin{equation}\label{2.6.2a}
R_{\mu\sigma}^{\;\;\;\;\lambda\sigma}e^{i}_\lambda
-\frac{1}{2}Re^i_{\mu}=\frac{1}{\varphi}\Sigma^i_\mu -
\frac{1}{2\varphi}U\/(\varphi)e^i_{\mu}
\end{equation}
\begin{equation}\label{2.6.2b}
\varphi\left( T^\alpha_{ts} - T^\sigma_{t\sigma}e^\alpha_s +
T^\sigma_{s\sigma}e^\alpha_t \right)
=\de{\varphi}/de{x^t}e^\alpha_s - \de{\varphi}/de{x^s}e^\alpha_t +
S^\alpha_{ts}
\end{equation}
while the Euler--Lagrange equation for the scalar field is given
by
\begin{equation}\label{2.6.2c}
R=U'\/(\varphi)
\end{equation}
\end{subequations}
Inserting eq. \eqref{2.6.2c} in the trace of  eqs. \eqref{2.6.2a},
we obtain an algebraic relation between the matter trace
$\Sigma\/$ and the scalar field $\varphi\/$ expressed as
\begin{equation}\label{2.6.3}
\Sigma -2U\/(\varphi) + \varphi\/U'\/(\varphi)=0
\end{equation}
Now, under the conditions ${\displaystyle
U\/(\varphi)=\frac{2}{\varphi}V\/(\varphi)\/}$, where
$V\/(\varphi)\/$ is defined as in eq. \eqref{2.4.7} and $f''\not
=0\/$. It is easily seen that the relation \eqref{2.6.3}
represents exactly the inverse of \eqref{2.4.6}. In fact, from the
definition of the effective potential \eqref{2.4.7} and the
expression $F^{-1}\/(X)=f'\/(X)X - 2f\/(X)\/$ we have
\begin{equation}\label{2.6.4}
U\/(\varphi) =\frac{2}{\varphi}V\/(\varphi)= \frac{1}{2}\left[
F^{-1}\/((f')^{-1}\/(\varphi)) +
\varphi(f')^{-1}\/(\varphi)\right]=
\left[\varphi(f')^{-1}\/(\varphi)
-f\/((f')^{-1}\/(\varphi))\right]
\end{equation}
so that
\begin{equation}\label{2.6.5}
U'\/(\varphi) = (f')^{-1}\/(\varphi) +
\frac{\varphi}{f''\/((f')^{-1}\/(\varphi))} -
\frac{\varphi}{f''\/((f')^{-1}\/(\varphi))} = (f')^{-1}\/(\varphi)
\end{equation}
and then
\begin{equation}\label{2.6.6}
\Sigma = - \varphi\/U'\/(\varphi) +2U\/(\varphi) =
f'\/((f')^{-1}\/(\varphi))\/(f')^{-1}\/(\varphi)
-2f\/((f')^{-1}\/(\varphi)) = F^{-1}\/((f')^{-1}\/(\varphi))
\end{equation}
In view of the latter relation,  eqs. \eqref{2.6.2} result to be
equivalent to eqs. \eqref{2.2.7} and \eqref{2.2.8}. This fact
proves the  equivalence between $f(R)\/$-gravity and scalar-tensor
theories with torsion  obtained also in the ${\cal J}$-bundles
framework.

\section{Conclusions}
In this paper, we have discussed the $f(R)$-theories of gravity
with torsion in the ${\cal J}$-bundles framework.

This formalism gives rise to a new geometric picture which allows
to put in evidence several features of the theories, in particular
their symmetries and conservation laws. In particular, due to the
fact that the components of the torsion and curvature tensors can
be chosen as fiber $\cal J$-coordinates on ${\cal
J}\/(\E\times\C)\/$, the field equations can be easily obtained in
suitable forms where the role of the geometry and the sources is
clearly defined.

Furthermore, such a representation allows to classify the
couplings with respect to the various matter fields whose global
effect is that to enlarge and characterize the $\cal J$-bundle.

We have given specific examples of couplings taking into account
Dirac fields, Yang-Mills fields and spin fluids. In every case,
the $\cal J$-vector fields allow to write the $f(R)$-field
equations in such a way that curvature, torsion and matter
components have a clear and distinct role into dynamics.

This result is particularly useful in cosmology where the role of
sources is crucial to define dynamics and, in some sense, to
coherently match the observations. In fact, the passage from the
"microscopic" domain of field equations and a suitable average
domain  for cosmology is extremely relevant to define
self-consistent cosmological models having well-founded
theoretical bases. For example, in the present state of cosmology,
one of the main shortcomings is due to the fact that a huge amount
of models explain the "same" data and, in our opinion, the
degeneracy could be due to the fact that the relation between the
microscopic and the macroscopic descriptions are often not well
defined, a part the very crucial issue to have homogeneous and
reliable data at any redshift. In this sense, giving a priori a
straightforward classification of curvature, torsion and matter
components is extremely useful.

In a forthcoming paper, we will show how all these considerations
could contribute to give a geometric, well-founded view of the
dark side of the universe.


\begin{thebibliography}{99}

\bibitem{CVB1} R.~Cianci, S.~Vignolo  and D.~Bruno,
The geometric framework for Yang--Mills theories, {\it J. Phys. A:
Math. Gen.\/} {\bf 36} (2003)  8341.

\bibitem{CVB2} R.~Cianci, S.~Vignolo and D.~Bruno,
Geometric aspects in Yang--Mills gauge theories, {\it J. Phys. A:
Math. Gen.\/} {\bf 37} (2004) 2519.

\bibitem{VC} S.~Vignolo and R.~Cianci,
 A new geometric look at gravity coupled with Yang--Mills fields,
 {\it J. Math. Phys\/},  {\bf 45\/} (2004) 4448.

 \bibitem{Ivanenko} D. Ivanenko and G. A. Sardanashvily,  The gauge treatment of gravity, {\it Phys.
Rep.} \textbf{94} (1983) 1.


\bibitem{CCSV1} S.~Capozziello, R.~Cianci, C.~Stornaiolo and S.~Vignolo,
$f(R)$-gravity with torsion: the metric--affine approach, \cqg
{\bf 24} (2007)  6417.

\bibitem{Copeland} E.J. Copeland, M. Sami, S. Tsujikawa, Dynamics of dark energy, \ijmp
{\bf D 15} (2006) 1753.

\bibitem{Odinojiri} S.Nojiri and S.D. Odintsov, Introduction to
modified gravity and gravitational alternative for dark energy,
{\it Int. J. Geom. Meth. Mod. Phys.} {\bf 4} (2007) 115.

\bibitem{GRGrew}
S. Capozziello and M. Francaviglia,   Extended Theories of Gravity
and their Cosmological and Astrophysical Applications, \grg {\it
Dark Energy: Special Issue} (2008) arXiv: 0706.1146 [astro-ph].

\bibitem{brans} C. Brans and R.H. Dicke, Mach's principle and
relativistic theory of gravitation, \pr {\bf 124} (1961) 925.

\bibitem{cimento} S. Capozziello, R. de Ritis, C. Rubano, and P.
Scudellaro, Noether Symmetries in Cosmology, {\it La Riv. del N.
Cimento} {\bf 4} (1996) 1.

\bibitem{sciama} D.W. Sciama, On the origin of inertia, \mnras {\bf 113} (1953) 34.

\bibitem{francaviglia} G. Magnano, M. Ferraris, and M.
Francaviglia,  Nonlinear gravitational Lagrangians, \grg {\bf 19},
(1987) 465 .

\bibitem{odintsov} I.L. Buchbinder, S.D. Odintsov, and I.L. Shapiro,
                    Effective Action in Quantum Gravity, IOP Publishing
                   (1992) Bristol.

\bibitem{birrell} N.D. Birrell and P.C.W. Davies, Quantum Fields
in Curved Space, Cambridge Univ. Press, Cambridge (1982).

\bibitem{vilkovisky} G. Vilkovisky, Effective action in Quantum
Gravity, \cqg {\bf 9} (1992) 895.

\bibitem{ottewill} J. Barrow and A.C. Ottewill, The stability of
general relativistic cosmological theory,  {\it J. Phys. A: Math.
Gen.} {\bf 16} (1983) 2757.

\bibitem{starobinsky} A.A. Starobinsky, A new type of isotropic
cosmological models without singularity, \pl {\bf B 91} (1980) 99.

\bibitem{kerner}  J.P. Duruisseau and R. Kerner, The effective
gravitational Lagrangian and the energy--momentum tensor in the
inflationary universe,  \cqg {\bf 3} (1986) 817.

\bibitem{noi}
S. Capozziello, Curvature Quintessence, {\it Int. J. Mod. Phys.} {\bf D 11} (2002) 483.\\
S.D. Odintsov, S. Nojiri, Where new gravitational physics comes from: M Theory?, \pl {\bf  B 576} (2003) 5\\
S.M. Carroll, V. Duvvuri, M. Trodden, M. Turner, Is cosmic
speed--up due to new gravitational physics, \pr {\bf D 70}
(2004) 043528.\\
G. Allemandi, A. Borowiec, M. Francaviglia, Accelerated
cosmological models in first order nonlinear gravity, \pr {\bf D
70} (2004) 103503.

\bibitem{ladek} S. Capozziello,  Dark Energy Models toward observational tests and data,
 {\it Int. Jou. of
Geom. Methods in Mod. Phys.} {\bf 4} (2007) 53.

\bibitem{mimick}
S.~Capozziello, V.~F.~Cardone, A.~Troisi, Reconciling dark energy
models with $f(R)$ theories of gravity, \pr  {\bf D 71} (2005)
043503.

\bibitem{magnano-soko}
 G. Magnano and L.M. Sokolowski, On physical equivalence between nonlinear gravity theories and a
 general relativistic self--gravitating scalar field, \pr  {\bf D 50} (1994) 5039.

\bibitem{faraoni}
V. Faraoni, Cosmology in Scalar-Tensor Gravity, Kluwer Academic,
Dordrecht (2004).

\bibitem{conformalCQG}
S. Capozziello, R. de Ritis, A.A. Marino,  Some aspects of the
cosmological conformal equivalence between "Jordan Frame" and
"Einstein Frame",
 \cqg {\bf 14} (1997) 3243.


\bibitem{palaeinstein} A. Einstein,  Metric--affine variational principles in
general relativity. I. Riemannian space-time, {\it Sitzungsber.
Preuss. Akad. Wiss.} (1925) 414.

\bibitem{frafe} M. Ferraris, M. Francaviglia, C. Reina, {\it J. Math. Phys.} {\bf
24} (1983) 120.

\bibitem{ACCF}
G. Allemandi, M. Capone, S. Capozziello, M. Francaviglia,
Conformal aspects of Palatini approach in Extended Theories of
Gravity, \grg {\bf 38} (2006) 33.

\bibitem{Hehl} F.~W.~Hehl and B.~K.~Datta, Nonlinear Spinor
Equation  and Asymmetric Connection in General Relativity, {\it J.
Math. Phys.\/} {\bf 12\/} (1971) 1334.

\bibitem{VCB2} S.~Vignolo, R.~Cianci and D.~Bruno,
On the Hamiltonian formulation of Yang--Mills gauge theories, {\it
Int. J. Geom. Methods Mod. Phys.\/} {\bf 2\/} (2005) 1115.

\bibitem{VCB1} S.~Vignolo, R.~Cianci and D.~Bruno,
A first--order purely--frame formulation of General Relativity,
\cqg {\bf 22\/} (2005)  4063.

\bibitem{VCB3} S.~Vignolo, R.~Cianci and D.~Bruno,
General Relativity as a constrained Gauge Theory, {\it Int. J.
Geom. Methods Mod . Phys.\/} {\bf 3\/} (2006) 1493.

\bibitem{VM} S.~Vignolo and E.~Massa,
A vielbein formulation of  unified Einstein--Maxwell theory, {\it
Class. Quantum Grav.\/} {\bf 23\/} (2006) 6781.


\bibitem{Hehl-Heyde-Kerlick} F.~W.~Hehl, P.~von der Heyde and G.~D.~Kerlick,
General Relativity with spin and torsion and its deviations from
Einstein's theory, {\it Phys. Rev. D.},  {\bf 10} (1974) 1066.

\bibitem{Ray-Smalley} J.~R.~Ray and L.~L.~Smalley,
Spinning fluids in the Einstein--Cartan theory, \pr {\bf D 27}
(1983) 1383.

\bibitem{Gasperini} M.~Gasperini, Spin--Dominated Inflation in the Eistein--Cartan Theory,
\prl  {\bf 56} (1986) 2873.

\bibitem{deRitis} R.~de Ritis, M.~Lavorgna, G.~Platania and C.~Stornaiolo,
Spin fluid in Einstein--Cartan theory: A variational principle and
an extension of the velocity potential representation, \pr {\bf D
28} (1984) 713.

\bibitem{German} G.~Germ\'an,
Brans--Dicke--type models with torsion, \pr {\bf D 32} (1985)
3307.

\bibitem{Sung} Sung--Won Kim,
Brans--Dicke theory in general space--time with torsion, \pr {\bf
D 34} (1986) 1011.


\end{thebibliography}
\end{document}